\documentclass[11pt,twoside]{article}
\usepackage{asp2010}

\resetcounters

\begin{document}

\title{The ADS All-Sky Survey}
\author{Alberto Pepe$^1$, Alyssa Goodman$^1$, and August Muench$^1$
\affil{$^1$Harvard-Smithsonian Center for Astrophysics}}

\begin{abstract}
The ADS All-Sky Survey (ADSASS) is an ongoing effort aimed at turning
the NASA Astrophysics Data System (ADS), widely known for its
unrivaled value as a literature resource for astronomers, into a data
resource. The ADS is not a data repository \textit{per se}, but it implicitly contains valuable holdings of astronomical
data, in the form of images, tables and object references contained
within articles. The objective of the ADSASS effort is to
extract these data and make them discoverable and available 
through existing data viewers. The resulting ADSASS data layer promises to greatly enhance
workﬂows and enable new research by tying astronomical literature and data assets
into one resource.  
\end{abstract}

\section{Introduction}
The Astrophysics Data System (ADS) is the premier literature resource for the astronomical
community. It maintains three bibliographic databases containing
roughly 9 million records, 4.5 million scanned pages, and 1.2 million
fulltext articles. Integrated in its databases, the ADS provides
access and pointers to a wealth of external resources, including
electronic articles, data catalogs and archives. While virtually all active
astronomers use the Astrophysics \textit{Data} System as a literature resource, very few know that
it was originally intended as a
\textit{data} resource \citep{ads_orig}. 

Nowadays, the ADS is not a data repository \textit{per se}, yet it implicitly contains valuable holdings of astronomical
data, in the form of images, tables and object references contained
within articles. The objective of the ADSASS effort is to
extract these data and make them discoverable and available 
through existing data viewers. The outcome of the ADSASS project consists of
a data layer, usable by any astronomical program, which documents,
links, and aggregates information about several astronomical data
pointers.

We focus on the extraction of three categories of data. First, we plan
to assemble astronomical object references provided and curated by
external databases and ``astrotag'' every article in the ADS. Similar
to \textit{geotags}---which are used to reference earth-based locations---
\textit{astrotags} are spatial and temporal annotations about celestial objects.
Second, we employ optical data, in the form of images extracted
from ADS articles to ``astroreference''
every image in the ADS fulltext corpus. Similar to \textit{georeferencing}---which refers to
the alignment of a map for overlay within a given earth-based
coordinate system--- \textit{astroreferencing} allows astrometric
alignment of an image of the sky according to given coordinates, orientation, and pixel scale.
Third, we inspect the metadata of non-optical images (such as text and image caption) to assign
programmatically a position or a source name. 

These data sources are then assembled to generate the ADS All-Sky
Survey layer (ADSASS) composed of an all-sky literature heatmap, and an
historical data layer. In this article, we present the technical implementation and proposed
use of the ADSASS.

\section{Technical implementation}
\begin{figure}[h!]
\centering
\includegraphics[width=1.0\textwidth]{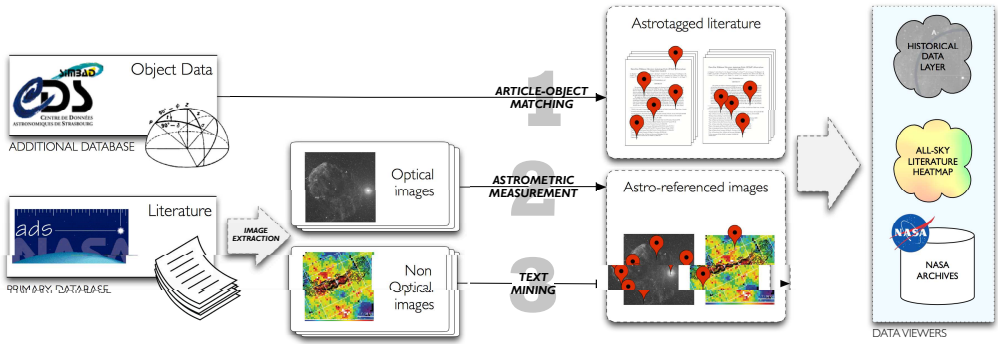}
\caption{The ADS All-Sky Survey: Technical implementation}
\label{fig:techplan} 
\end{figure}
 
The procedures, data sources, and outcomes that are part of the
ADS All-Sky Survey project are depicted, in a summarized
format, in Figure \ref{fig:techplan}. The principal data sources are the
NASA ADS bibliographic database, and the SIMBAD
object catalogue, maintained by the Centre de Données astronomiques de
Strasbourg (CDS).  

The first step of the astrotagging procedure involves using all the
holdings of the ADS library for which object references exist to populate a
database of astrotagged literature. This mechanism of article-object
matching generates an initial database that associates ADS articles with their manually curated
object references provided by CDS. The purpose of this first step is to aggregate and publish the
conceptual linkages that already exist between ADS and CDS
holdings. The outcome of this article-object matching procedure is a database of
astrotagged literature. For example, in this procedure a paper by astronomer John
Huchra which discusses the \texttt{M31} \citep{1987AJ.....93..779H} would be
astrotagged with the following elements: a) the referenced objects, each identified by their SIMBAD reference name
(e.g., M 31), and co-ordinates (e.g., 00 42 44.330, +41 16
07.50), and b) a timestamp, identified as the year of publication of the
  article (e.g., 1987). However, this article-object matching procedure is only possible for
the portion of the ADS database for which an object
categorization exists. In steps 2 and 3, we
attempt to annotate those papers/images for which no classification is available by automatically inspecting articles for object and/or
coordinate information. 

As shown in Figure \ref{fig:techplan}, we plan to extract images from all fulltext
articles in the ADS. For the
purpose of this project, extracted images of interest fall in two
categories: \textit{optical images} and \textit{non-optical images}.

\begin{figure}[h!]
\centering
\includegraphics[width=1.0\textwidth]{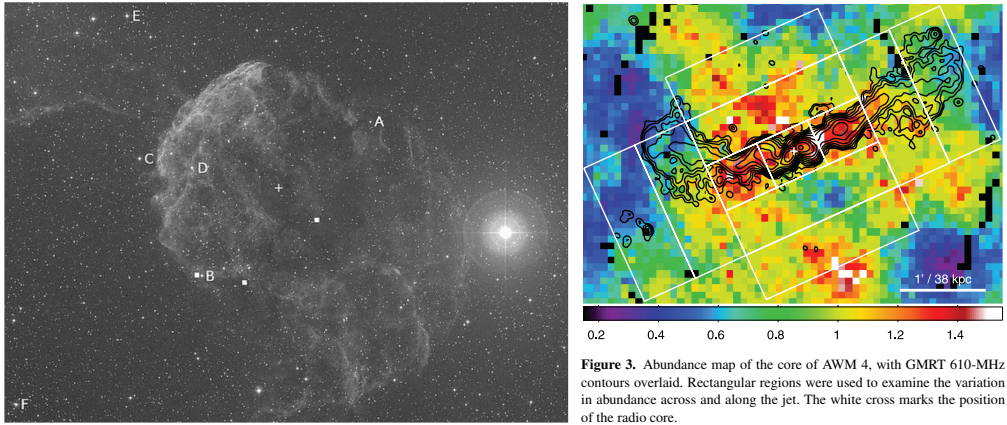}
\caption{(a) On the left: a sample optical image from Indriolo et
  al. (2010) ---     (b): On the right: a sample non-optical image from O' Sullivan et al. (2011)}
\label{opticalnonoptical} 
\end{figure}

A sample extracted optical image is presented in the left portion of
Figure \ref{opticalnonoptical} (a). The image
was extracted from a recent article by
\cite{2010ApJ...724.1357I}. When fed to \url{astrometry.net}
\citep{lang2010}, this image is correctly resolved as: 

\begin{footnotesize}
\begin{quote}
RA,Dec center: (06:17:4.161, +22:33:23.918); Orientation:  -179.77 deg
E of N; Pixel scale: 4.56 arcsec/pixel; Field size: 77.89 x 58.19
arcminutes; Field contains: Propus (ηGem), IC 443.
\end{quote}
\end{footnotesize}

This reference information (supernova remnant IC443 and the triple-star
system Propus=ηGem) would become part of the astrotags attached to this article,
in addition to a time stamp given by publication date.  In this case,
however, in addition to astrotagging the article, we would also
astroreference the image, using the coordinates, orientation, and
pixel scale returned by \url{astrometry.net}. In this way, a database of
literature-extracted astroreferenced images is populated (step 2 of Figure \ref{fig:techplan}).

When astrometric measurement fails, enough textual metadata may be
available to assign either a position or a source name, and optionally
a platescale and waveband to the images. An example of non-optical image is a metal abundance map from a
recent paper by \cite{2011MNRAS.411.1833O}, shown in Figure
\ref{opticalnonoptical} (b). This image cannot be calibrated
astrometrically, yet, the figure caption gives enough information
to extract the needed source name and wavelength metadata. In this
case, it is clear from the caption that the source depicted in this
cluster is the AWM 4 cluster of galaxies.  SIMBAD or NED can tell us
that AWM4 is also known as RXC J1604.9+23559, with coordinates RA: 16
04 57.0, DEC: +23 55 14. Thus, using figure caption metadata, this
article/figure would be astrotagged with name and coordinate
information relative to the aforementioned catalog objects (step 3 of Figure \ref{fig:techplan}).

\section{The ADS All-Sky Survey in action} 
The ADSASS will yield two products: an \textit{all-sky literature
  heatmap}, built from the astrotagged literature database, showing what parts of the sky have been
written about in the literature, and in what contexts; and an
\textit{historical data layer}, built from the astroreferenced image database, offering literature-extracted images for
analysis and overlay on contextual images. Thanks to the proliferation of virtual observatory data
viewing systems over the past few years, the options for visualizing
these layers are many, and they include compatible all-sky viewers
such as the WorldWide Telescope (Microsoft), Aladin (CDS), Google Sky (Google), and MASTview10 (NASA).

This composite ADSASS layer promises to enable and enhance astronomical research
in many ways. First, the ADSASS will draw on the constantly-updated full inventory of
astrotagged ADS literature, to enable on-demand faceted \textit{visual knowledge
discovery} for any target or
topic. For example, a graduate student will be able to generate on the fly a
heatmap of the Orion molecular cloud to figure out which parts of the cloud have existing
molecular line data, where they are missing, and where specific
science topics (e.g., jets and outflows)  have been explored in
published works.  

Second, the faceted heat map of the ADSASS will enable new forms of
\textit{data discovery}. Consider the problem of discovering published data on ``the radial
velocities of young stars'' in a particular region of interest.  A
purely text-based literature search using those words would miss
targets of interest only mentioned in the literature as, for example, ``host stars'' in
``exoplanet surveys''. Instead of this text-only search, exploring the faceted
heat map of the ADSASS, researchers will be able to traverse related
data and objects using a visualization-aided literature search.
 
Third, researchers’ investigations of the ADSASS historical data layer will
facilitate the \textit{study of astrophysical events}, thanks to its synoptic image view.
Eruptive events, such as an accretion driven outburst in a very young star, reveal objects
that were previously invisible. Synoptic monitoring surveys such as
the Palomar Transient Factory are designed to identify these eruptive
events but on relatively short timescales. The historical data layer will capture and expose data that was
never saved as such in the first place, extending the time baseline
for many parts of the sky back almost 100 years (much longer than the
epoch provided by the Digital Sky Surveys), and providing a framework
to identify solitary or recurrent astrophysical events. 

\section{Conclusion}
The ADS All-Sky Survey reunites data and literature holdings
into a powerful astronomical research tool. The mission of the ADSASS
is to ``liberate'' data sources that are contained in scientific papers
as images, tables, object references, and metadata. The
publication of these data in the form of a faceted heat map of the sky
and a historical data layer will enable new forms of astronomical
research and ultimately augment the intrinsic value of the scientific
papers in which those data were embedded.

\acknowledgements The ADS All-Sky Survey was selected for funding as
part of the NASA Research Opportunities in Space and Earth Sciences
(ROSES-2011) for the Astrophysics Data Analysis Program (ADAP). The
authors would like to thank collaborators and project supporters
T. Boch of CDS, A. Accomazzi of ADS, D. Hogg of NYU,
J. Fay of Microsoft Research, and A. Conti of STScI.

\bibliographystyle{asp2010}
\bibliography{O23}

\end{document}